\documentstyle[11pt,leqno,fleqn]{article}
\input{psfig.tex}
\title{Comments on "Oriental magic mirrors and the Laplacian image"}
\author{A. Kwang-Hua Chu} 
\date{Department of Physics, Xinjiang University, Urumqi 830046, PR China}
\begin{document}           
\maketitle                 
\begin{abstract}
We make some remarks on Berry's paper [{\it Eur. J. Phys. } 27 (2006) 109-118]. \newline

\end{abstract}
\doublerulesep=6mm        
\baselineskip=6mm
\oddsidemargin+1mm         
\bibliographystyle{plain}
Berry just presented a comprehensive explanation of the
physical part for the optics of the {\it magic} mirror [1]. The story related to this
strange metal mirror could be traced in [2-4]. The characteristic is :
the raised images appeared in a screen are on the opposite side of an
opaque mirror but not the reflecting face [5-6]! This amazing optical effect for many centuries made a tremendous impression on observers, from which
such mirrors became known as magical. Master craftsmen
passed down the secret of their fabrication from generation
to generation. The ancients were unable to give the phenomenon
a convincing explanation, but it was reflected in the old
saying "the truth always comes out in the sun." [5-6].
\newline
In fact, archeological finds contain extremely scanty information
on the magic mirrors of China. One of the first such
mirrors was discovered during excavations of the tomb of the
exalted grandee Yu in the province of Uhan in the south of
China and has been dated to approximately 500 B.C. These
years belong to the Han period of history, when, in the fabrication
of mirrors, special significance was attached to the
purity of the metal. The necessity for such careful fabrication
of the mirrors was caused by ritual-magical purposes; these
were often reflected in inscriptions and images on the mirrors,
which were supposed to keep evil forces away from
their owners (cf. Fig. 2 in [6])!. Faith in the power and magical properties
of these symbols in Han China was very widespread. Other legend is : the reverse side of the mirrors bore raised images and
inscriptions of lyrical, symbolic, and edifying content: "a
wise man uses his mind like a mirror" [6].
\newline
In China, attempts were made from ancient times to explain
the production of the remarkable optical effect created
by magic mirrors. The first version that has come down to us
apparently belongs to the scholar  Kua Shen (1031-1095)! [4].
He explained this phenomenon by saying that, in casting, the
thinner part cools faster than the thick part, and this causes
the formation of small deformations of the profile, not observable
by the naked eye. At the same time, he pointed out
that very thin mirrors did not manifest this property, and
gave credit to the high skill of the ancients. At the end of the
thirteenth century, Dzhou Mi observed a mirror that reproduced
in reflected light the finest details of the image on the
back side. \newline
The archeologist Uchkhi Ien gives another explanation
of the effect, associating it with the use of bimetals of different
density. Thus, if an image of a dragon was created on
the back side by casting, the image was engraved on the
front side and was then filled with bronze of another density,
after which the surface was carefully polished. He saw such
a mirror that had been cut apart and was absolutely certain
that his explanation was correct [6]. Other attempts could be traced in [1-2].
\newline
In such an interesting regime of geometrical optics, the image intensity
could be given simply by the Laplacian of the height function of the relief as Berry demonstrated in [1]. For instance,
Berry used the error function (cf. Eqn. (14) in [1]) to model smoothly and approximately  the single step (the $l$-smoothed step, with height $h_0$). The present author, however, likes to propose another way :
use $Tanh$ (hyperbolic tangent function) to model the sharp step.
It reads
\begin{equation}
 h(x)=C_0 h_0 \frac{e^{x/l_t} - e^{-x/l_t}}{e^{x/l_t} + e^{-x/l_t}}, \hspace*{12mm}
 l_t=a_0 l,
\end{equation}
$C_0$ is a normalization constant and $a_0$ is a constant for adjusting the sharpness.
The result and comparison is shown in Figure 1. There is no doubt that our proposal
could be either smooth (enough for the Laplace operator) or sharp enough (to approximate the step). The obtained function could be easily implemeted in
\begin{displaymath}
 I_{Laplacian} =1 + Z \nabla^2 h({\bf r}), \hspace*{12mm} Z=2 D/M,
\end{displaymath}
where $D$ is the distance of the scrreen from the reference plane,
$M$ is the magnification, $Z$ is the reduced distance, and ${\bf r}$ is the demagnified
observation position [1]. \newline
Meanwhile, as commented by Berry in [1] : "It is possible that there are different types of magic mirror, where for example the relief is etched directly  onto the reflecting surface and protected by a transparent film [7], but these do not seem to be common. Sometimes, the pattern reflected onto a screen is different from that on the back, but this is probably a trick, achieved by attaching a second layer of bronze, differently embossed, to the back of the mirror.". Berry only briefly discussed the manner in which the pattern
embossed on the back gets reproduced on the front at the end of [1] : "Referring to (11), this involves the sign of the coefficient $a$ in the relation between $h_{back}$ and $h$. There have been several speculations about the formation of the relief. One is that the relief is generated  while the mirror is cooling, by unequal contraction of the thick and thin parts of the pattern [8]; it is not clear what sign of $a$ this leads to. Another [9] is that cooling generates stresses, and that during vigorous grinding and polishing the thin parts yield more than the thick parts, leading to the thick parts being worn down more; this leads to $a<0$. However, this seems to contradict the observations, which point firmly to
$a > 0$ : bright (dark) lines on the image, indicating low (high) sides of the steps on the reflecting face, are associated with the low (high) sides of the steps on the back (cf. figure 7(a) in [1]), not the reverse (cf. figure 7(b) in [1]).". Here, the present author would like to mention the significance of the material which made the strange mirror. \newline
As reported in [6], a mirror of the same size and thickness as the Han
mirror was fabricated in the University of modern China. They used a mirror from the Shanghai museum as a
model for the casting. The copy was fabricated from an alloy
containing 73\% copper, 23\% tin, and 4\% lead. After cooling,
the mirror was ground to a thickness of 0.5mm in the thin
places. When the copy was illuminated, it behaved exactly
like the original. This could be compared to those larger magic mirrors, now found in various collections, belong to the period of government of the Ming
dynasty (1368-1644). They are distinguished by the size and
character of the images. The reflective surface was made
somewhat convex and was carefully polished by means of a
mercury amalgam. The back side often had intricate images
of birds, flowers, or dragons or scenes from mythology. The
spread in height of the relief is about 25\%. The production
technique was casting, using the lost-wax technique. One of the largest of such mirrors, fabricated later in China, in 1875, is 52 cm in diameter and 1.3 cm thick and
weighs more than 12 kg. As with other Chinese mirrors, it is
fabricated from bronze, being an alloy of copper (80\%), tin
(15\%), and lead (5\%) [6]. Finally, from all previous information, it is evident that, in the image
obtained on a screen, dark zones appear where the light is
deflected by convex microsections of the surface corresponding
to the thinner regions of the mirror, while bright zones
are formed by flat microsections corresponding to the thicker
regions. Parts of the latter  were illustrated in [1] clearly.

\newpage

\psfig{file=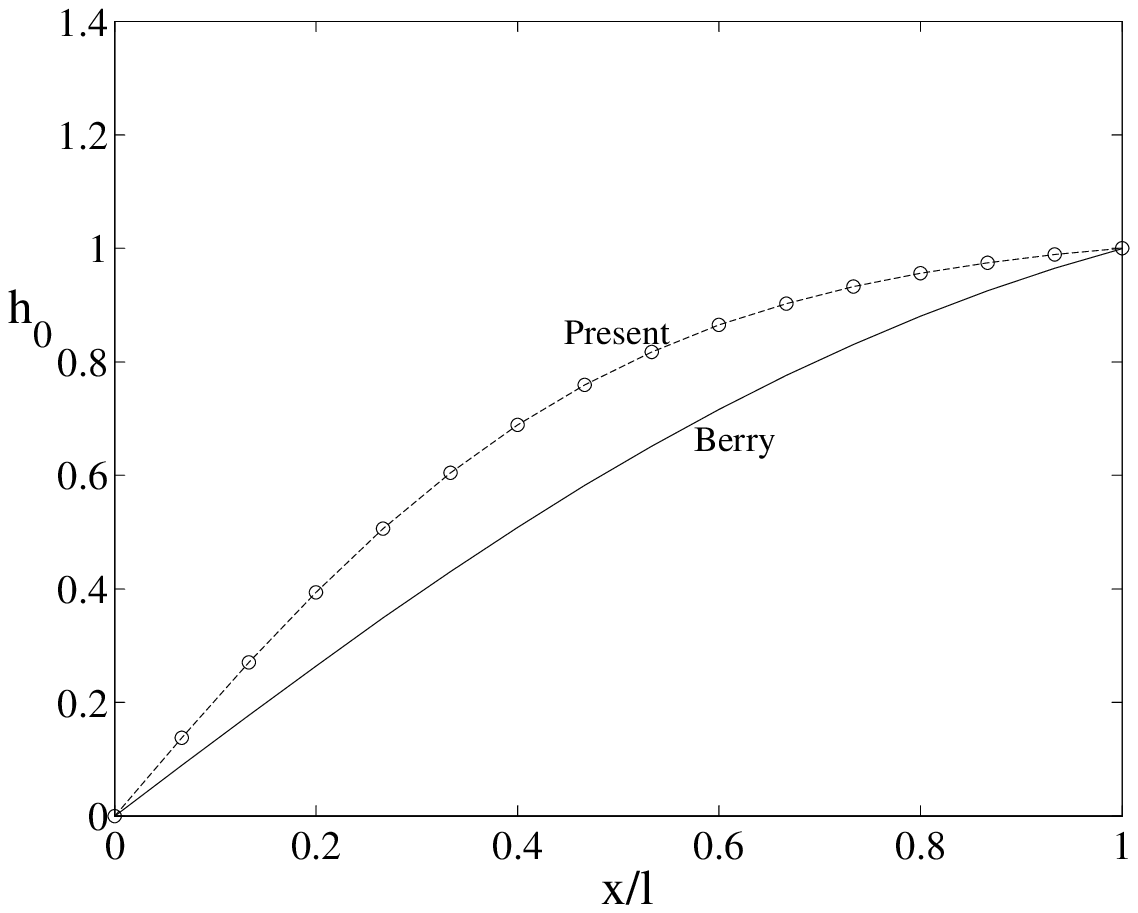,bbllx=0.1cm,bblly=13.8cm,bburx=15cm,bbury=23.8cm,rheight=9.6cm,rwidth=9.6cm,clip=}
\vspace{2mm}
\begin{figure}[h]
\hspace*{6mm} Fig. 1 \hspace*{1mm} Comparison of the smoothing of
the single step : the $l$-smoothed step, with \newline  \hspace*{6mm}
height $h_0$ (cf. Berry's proposal [1]).
 We adopt the hyperbolic tangent  (Tanh) function \newline  \hspace*{6mm}
 and the result is much more close to
 the sharp step compared to that \newline  \hspace*{6mm} by Berry (cf. Eqn. (14) in [1]).
\end{figure}

\begin{thebibliography}{99}
\doublerulesep=2mm        
\baselineskip=3mm
\bibitem{S:Plate}  Berry MV 2006 Oriental magic mirrors and the
Laplacian image {\it Eur. J. Phys. } {\bf 27}  109-118.
\bibitem{M:2001} Auckland G 2001 Magic Mirrors, or Through the Looking Glass http://www.grand-illusions.com/magicmirror/magmir1.htm.
\bibitem{1975:M} Lubo-Lesnichenko EI 1975 Imported Mirrors of the Minusinsk Basin (Glavn.
Red. Vostochn. Lit., Moscow, 1975).
\bibitem{UK:M}  Needham J and  Ling W 1962 Science and Civilisation in China (Cambridge,
1962), vol. 4, pp. 94-97.
\bibitem{J:Opt-Tech}  Seregin DA,  Seregin AG and Tomilin MG  2004
Method of shaping the front profile of metallic mirrors with a given relief of its back
surface {\it J. Opt. Technol.} {\bf 71} 121-122.
%
\bibitem{M:M}  Saines G and  Tomilin MG  1999  Magic mirrors of the Orient
{\it J. Opt. Technol.}
{\bf 66} 758-765.
\bibitem{Mak:2001} Mak S-Y and Yip D-Y 2001 Secrets of the Chinese magic mirror replica {\it Phys. Educ.} {\bf 36} 102-107.
\bibitem{Yan:1992} Yan Y-L 1992 Three demonstrations from ancient Chinese bronzeware {\it Phys. Teach.}  {\bf 30} 341-3.
\end{thebibliography}
\end{document}